\documentclass[twocolumn]{aa}
\usepackage{graphicx}
\usepackage{txfonts}
\usepackage{natbib}
\bibpunct{(}{)}{;}{a}{}{,}
\newcommand{\pc}{{\rm pc}}
\newcommand{\kpc}{{\rm kpc}}
\newcommand{\mas}{{\rm mas}}
\newcommand{\Msun}{{\rm M_\odot}}
\newcommand{\Rg}{\mbox{$\rm r_g$}}
\newcommand{\Rl}{\rm r_0}
\newcommand{\jet}{\rm _j}
\newcommand{\disk}{\rm _d}

\begin{document}
   \title{Magnetic collimation of the relativistic jet in M87}

   \author{J. Gracia\inst{1}
	  \and
	  K. Tsinganos\inst{1}
          \and
          S.~V. Bogovalov\inst{2}
          }

   \offprints{J. Gracia, \email{jgracia@phys.uoa.gr}}

   \institute{IASA and Section of Astrophysics, Astronomy and Mechanics,
     Department of Physics, University of Athens,\\ 
     Panepistimiopolis, GR--157\,84 Zografos, Athens, Greece
     \and
     Moscow Engineering Physics Institute (State University), 
     Kashirskoje shosse 31, 115409 Moscow, Russia
   }

   \date{Received; accepted}

   \abstract{ We apply a two-zone MHD model to the jet of M87. The
     model consists of an inner relativistic outflow, which is
     surrounded by a non-relativistic outer disk-wind. The outer
     disk-wind collimates very well through magnetic self-collimation
     and confines the inner relativistic jet into a narrow region
     around the rotation axis. Further, we show by example, that such
     models reproduce very accurately the observed opening angle of
     the M87 jet over a large range from the kiloparsec scale down to
     the sub-parsec scale.
   }

   \maketitle
%

\section{Introduction}
One of the oldest \citep{Curtis} and best studied extragalactic
jets is harboured by M87 in the Virgo cluster. This classical object
presents a nice opportunity to test a specific mechanism for jet
formation in some detail, namely magnetic self-collimation, first
pioneered by \citet{BP82}. Later, a overwhelming wealth of theoretical
investigations build upon this model. 

At the distance of M87 ($z=0.004$, \citet{Jacoby+90}) a miliarcsecond
of angular scale corresponds to a linear scale of 0.072 \pc. The
central mass of the AGN is approximately $3 \times 10^9 \, \Msun$
\citep{Macchetto+97}, which translates into a Schwarzschild radius
\Rg{} of about 0.0003 \pc{} or 0.0041 \mas. Thus at the arcsecond
scale, the jet of M87 has a length of 2 \kpc. The jet and its hot
spots have been systematically studied across the electromagnetic
spectrum from the radio to X-rays, both, with ground-based
observations and from satellites \citep[for a review see
e.g.][]{Biretta96}. The jet is clearly detected at mm wavelength with
a resolution of 0.009 \pc{} (30 \Rg) out to distances of about 2
\mas{} (500 \Rg) from the core. The initial opening angle is
approximately 60\degr{} on scales of about 0.04 \pc{} (100 \Rg) and
decreases rapidly until reaching 10\degr{} at a distance of 4 \pc{}
from the core \citep{Biretta+02}.

These observations suggest, that the jet of M87 is rather slowly
collimated across a length of several parsec (several $10^5$
\Rg). This scale is significantly larger than the radius of the black
hole or any of the characteristic orbits, e.g. the last-stable orbit
at 6 \Rg, but well below the size of the accretion disk, which can be
as large as 20 \pc{}. Therefore \citet{Biretta+02} conclude that the
accretion disk plays an important role in the initial jet
collimation. On the other hand, thermal X-ray observations
\citep{White+88} show that the jet is overpressured, 
since it is surrounded by gas of a thermal pressure, which is 10--20
times lower than the minimum pressure value required for the
synchrotron emission of the jet \citep{OHC89}. In addition, the
projected magnetic field vectors suggest a significant azimuthal
component of the magnetic field which can assist in confining the
jet. In this work, we provide a specific mechanism which illustrates
the assumption that the disk plays a pivotal role in magnetically
collimating the jet.

The prevailing view in jet formation theory is that relativistic and
non-relativistic jets are magnetically collimated, a view supported by
observational evidence \citep{Gabuzda+04}. The toroidal magnetic field
generated by the rotation of the jet's base, i.e. the underlying
accretion disk, spontaneously collimates part of the outflow into a
narrow region around the axis of rotation, as shown by numerous
analytical models, e.g. \citet{Lovelace76, Blandford76, BR76, BP82,
HN89, HN03, CLB91, ST94, VT98, VT99, V00}. Nevertheless,
to calculate the fraction of the collimated mass- and magnetic fluxes,
the possible formation of shock waves or the opening angle of the
outflow, one needs direct numerical simulation for every specific case
\citep[][and others]{KLB99, Ustyugova+99, KMS98, GVT05}.

In \citet{BT99} and \citet{TB00} it was found that, for a uniformly
rotating base, only a small part of the total mass- and magnetic flux
is collimated cylindrically. This fraction is only about 1\% of the
corresponding values of an assumed uncollimated outflow, i.e. before
rotation of the base sets in. Recently, this conclusion was confirmed
by \citet{KLB03} for outflows originating from an accretion
disk. However, observations and theoretical arguments indicate that a
higher fraction of mass- and magnetic flux should be collimated inside
the jet. A second difficulty of the original magnetic collimation
picture is that for relativistic outflows, the degree of collimation
of an initially radial wind is extremely small due to the
decollimating effect of the electric field and the large effective
inertia of the relativistic plasma \citep{B01}.

In a series of recent papers \citet{TB02,TB05} adopted a simple model
to demonstrate, that the mechanism of magnetically self-collimation of
outflows may also efficiently collimate even relativistic outflows,
provided that the system consists of {\em two components}: an {\em
inner relativistic plasma} originating from regions close to the
central source and an {\em outer non-relativistic wind} originating
from, e.g., the surrounding accretion disk \citep[see
also][]{SPA89}. In the particular case studied in \citet{TB02}, the
toroidal magnetic field in the inner relativistic outflow was
negligible by assuming that the angular velocity at its base is
negligible. Under such conditions the disc-wind plays the role of the
collimator and confines all the relativistic outflow into a collimated
fiducial jet around the axis. Steady state solution for such
relativistic jets were obtained with Lorentz factors up to
$\Gamma=5$. We stress, that all the magnetic and mass flux at the base
of the relativistic plasma is collimated into the relativistic jet.

We apply a similar model to the case of the jet of M87 to
demonstrate, that the magnetic confinement of an inner relativistic
outflow by a non-relativistic disk-wind may lead to a slow collimation
of the relativistic jet which fits the observations of the opening
angle. In the following sections we describe in more detail the model
and the numerical method used, apply this to the parameters
appropriate to the jet of M87 and present some results.

\section{Model and numerical method}
\begin{figure}
  \centering 
  \includegraphics[height=\columnwidth]{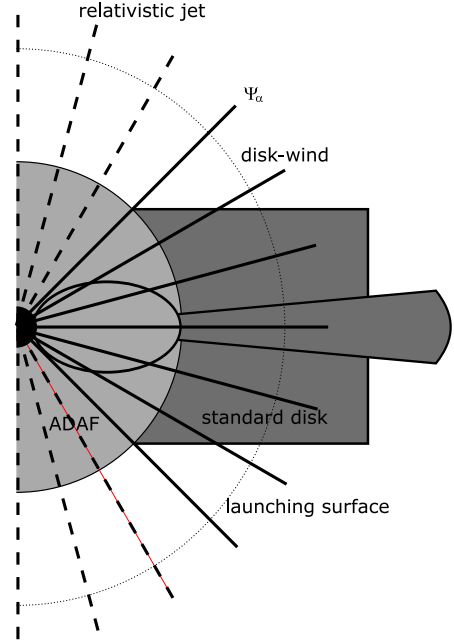}
  \caption{Illustrative sketch of the model (see text).}
  \label{fig:model}
\end{figure}

We adopt as a model an axisymmetric outflow consisting
of two components, which are implemented as boundary conditions. An
inner relativistic outflow, which originates from close to the central
black hole, and an outer non-relativistic one, originating from a
region which is no longer dominated by relativistic dynamics. This may
be realized by a similar two-component structure of the underlying
accretion flow consisting of an outer standard disk
\citep{SS73} and an inner hot flow, which can be either an advection
dominated accretion flow \citep{NY94, GPKC03} or the final plunging
region near the black hole, where relativistic dynamics dominates
through e.g. frame-dragging or the \citet{BZ77} effect. We assume that
initially radial outflows originate from both regions. In the
following we will refer to them as relativistic jet or outflow and
(non-relativistic) disk-wind, respectively. Figure \ref{fig:model}
shows a cartoon of the model.

The relevant physical properties of these two components are assumed
to be given at a spherical launching surface at a distance $\Rl$. We
further assume, that most physical quantities have constant values on
the launching surface within the two regions, respectively. The only exception
is the angular velocity as explained later. The boundary between the
two components is given by an angle $\alpha$ measured from the
rotation axis. The indices $\jet$ and $\disk$ designate quantities in
the jet and disk-wind component, respectively. The launching surface
is threaded by a radial magnetic field of strength $B_0$. We normalise
the magnetic flux $\Psi$ to $\Psi=0$ on the rotation axis and $\Psi=1$
on the equatorial plane.  The two outflow components are characterised
by their temperature in units of the rest-mass energy $T$ and the
radial outflow velocity in terms of the bulk Lorentz factor $\Gamma$.

For simplicity, the angular velocity $\omega$ in the
inner relativistic outflow is constant at a value
$\omega\jet$. In the disk-wind it drops smoothly to zero as
given by 
\begin{equation}
  \label{eq:angvel}
  \omega\disk(\Psi) = \omega\jet \left (\frac{\Psi_\alpha}{\Psi}
  \right)^2, 
\end{equation}
where $\Psi_\alpha = 1 - \cos \alpha$ is the magnetic flux at the
boundary between the two components. Further, for simplicity the
disk-wind is assumed to be dynamically cold, i.e. $T_d = 0$ and the
magnetic field strength at the launching surface to be fixed at $B_0 =
1 {\rm G}$ throughout this paper. This choice leaves us with the set
of adjustable free parameters $(\alpha, T\jet, \Gamma\jet, \omega\jet,
\Gamma\disk)$ and additionally the radius of the launching surface, which
is constraint by the numerical method that we use to solve the system
of equations.

We apply the same numerical method as \citet{TB02}. MHD problems in
general can be decomposed into two mathematical regimes, hyperbolic
and elliptical. In the hyperbolic regime, the problem can be treated
as an initial value Cauchy-type problem, i.e. given the conditions on
a specific surface as initial values, the steady state solution of the
flow further downstream can be calculated by direct integration. This
is numerically much easier than treating the problem in the elliptical
regime. If the local poloidal velocity of the flow exceeds the local
speed of the fast-mode Alfv\'en waves, then the flow is
hyperbolic. Therefore, we place the launching surface beyond the fast
surface and calculate the physical quantities along the flow by means
of the conserved MHD integrals of the flow.

Further, the magnetic field structure is solved self-consistently by
means of the transfield equation instead of assuming a given external
magnetic field. In this way, we obtain a self-consistent solution of
the full, steady-state, relativistic MHD equations, which
depends only on the boundary conditions assumed on the
launching surface. The latter must necessarily be located
beyond the fast surface of the outflow. More specifically, we do not
solve the problem inside the launching surface, and a self-consistent
solution of the global problem, i.e. including the sub-fast
region inside the launching surface, might show, that, in particular,
the assumed radial magnetic field inside the launching surface might
not be realized.

\section{Data basis and fitting results}
\begin{figure}
  \centering 
  \includegraphics[width=\columnwidth]{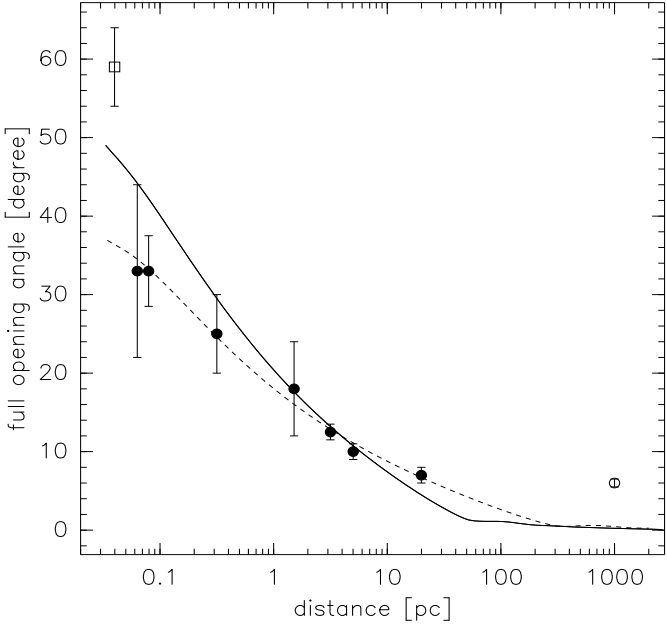}
  \caption{Comparison of the opening angle calculated from model A and
  model B, and the observational data for M87. The {\em black lines}
  represents model A ({\em solid}) and model B ({\em
  dashed}),respectively, while various symbols represent
  observational measurements. The data points marked by {\em black
  circles} was taken into account in the fitting procedure for model A
  and B. Model A fits, additionally, the innermost data point ({\em
  open square}).  The outermost measurement {\em open circle} was
  disregarded for both models, since it is located beyond knot A (see
  text).}
  \label{fig:modelA_B}
\end{figure}

The model has been applied to the jet of M87. In particular, we aimed
to reproduce the measured opening angle as collected and reported by
\citet{Biretta+02}. In observations the opening angle of the jet is
given by the width of the jet, which in turn is the length scale
beyond which the brightness of the jet drops of sharply. Our model
does not include any radiative processes and does not have any notion
of brightness. Instead, we define the boundary of the jet by a
specific magnetic flux line $\Psi_\alpha$ which separates the two
components at the launching surface. Everything inside this flux line
is assumed to be visible as jet, everything outside is not.

The observations extend from very small length scales of 0.04 \pc{} to
much larger scales of 1 \kpc{}. The largest scale data point is
actually obtained in regions downstream of the very prominent {\em
knot A} of the jet, which is located at a distance of
12\arcsec{} (865 \pc) from the core. It is generally believed, that
the jet outflow dramatically changes character at knot A, either
through interaction with the environment or through internal
micro-physics. We do not model either of these and therefore exclude
any data beyond knot A from our fitting procedure.

Similarly, we might exclude the innermost data point at 0.04 \pc. This
point is very close to the adopted launching surface at $\Rl = 10^{17}
{\rm cm}$. As discussed previously, we expect the largest
uncertainties in our model close to this surface. In particular, the
magneto-centrifugal mechanism is, under generic conditions,
most efficient in collimating the poloidal magnetic field near the
Alfv\'enic surface, which is located inside the fast surface. So, in
reality, the outflow will collimate even more efficiently near its
base, than can be calculated from our current model. To meet the
constraint, that the fast surface is within the launching surface, we
fix the magnetic field strength to $B_0 = 1 {\rm G}$.

We present two exemplary models: model A includes the smallest scale
in the data set, while the model B does not. Model A fits all data
inside knot A, as shown in figure \ref{fig:modelA_B}. The model
parameters are $(\alpha, T\jet, \Gamma\jet, \omega\jet,
\Gamma\disk)_{\rm A} = (25\degr, 2.5\,mc^2, 1.8, 2.7 \, \Rg/c,
1.02)$. The outflow is only moderately relativistic, both in terms of
its internal energy and in terms of its bulk motion with a Lorentz
factor of only $\Gamma\jet = 1.8$ or physical velocity $v\jet =
0.83\,c$. A better fit would be obtained if the curve were steeper at
the beginning and flattened out at larger distances.  Unfortunely, our
parameter studies showed, that it is very difficult to obtain such a
curve, i.e. to change its curvature. Basically, there are only three
degrees of freedom: displacing the curve in the vertical and
horizontal direction, and rotation around a pivotal point near the
origin. For this reason, the fit to the data points at intermediate
distances can be further improved only, if the data point at the
lowest distance is not taken into account.

This constraint leads us to model B, which fits all data inside knot A
excluding the innermost data point, as shown in figure
\ref{fig:modelA_B}. The model parameters are $(\alpha, T\jet,
\Gamma\jet, \omega\jet, \Gamma\disk)_{\rm B} = (19\degr, 2.5\,mc^2,
2.6, 2.7 \, \Rg/c, 1.02)$. The initial opening angle is considerably
smaller than for model A. This implies a higher magnetic field
strength at the boundary between the two zones, which leads to more
efficient collimation in the disk-wind and confinement of the
relativistic jet. To achieve large opening angles further down the
flow, this has to be compensated by a larger effective inertia. This
is accomplished through a higher initial poloidal velocity of Lorentz
factor $\Gamma\jet = 2.6$, i.e. physical velocity $v\jet = 0.92\,c$. We
note however, that a similar effect could also have been accomplished
with a lower collimation efficiency of the disk-wind by reducing the
angular velocity at its base, or even by increasing the effective
inertia through higher internal energy $T\jet$.

\section{Conclusions}

The two-zone model presented in this work fits nicely the observed
opening angle of the jet of M87 over at least three decades in linear
length scale.  It solves two problems of one-zone, magnetic,
self-collimation models by sharing the work between the two zones. The
inner relativistic zone provides the moderately high Lorentz factors
observed in a number of objects, but has a very low collimation
efficiency and would thus not result in a cylindrical or conical jet
of its own. This job is done by the outer non-relativistic
disk-wind. This zone collimates very efficiently and confines the
inner relativistic jet to a narrow region around the rotation axis. On
the other hand, by itself, it could not attain the necessary Lorentz
factors due to its non-relativistic nature.

The single most important ingredient seems to be the balance between
effective inertia of the relativistic outflow and the confining force
of the outer disk-wind. This equilibrium can be shifted in a number of
ways, e.g. towards higher effective inertia by increasing the outflow
velocity or internal energy of the relativistic jet or by decreasing
the collimation efficiency in the outer disk-wind by reducing the
angular velocity. In this way it is possible to produce similar
opening angle profiles with a number of different parameter sets. 

It is therefore very important to constrain observationally some of
the free parameters of the model. One obvious working point is the
location of the fast surface, i.e. Alfv\'enic and bulk velocity near
the base, which tells us where to start applying this model.
We note further, that \citet{Biretta+02} were never truly able to
pinpoint the location of the central object. Hence, their opening
angle determinations, even if impressive, are only tentative. This is
especially true for the innermost data point. If the central source
were further away from the jet, this would strongly decrease the
opening angle. Still such a opening angle profile could be easily
satisfied by parameters along the lines of model B.

Our model relies on the existence of of two dynamically
distinct regions. To communicate ideas, we identified them with an
inner ADAF or final plunging region of the flow, and a standard
disk. The implementation of this two zones as regions of constant
physical quantities, each, is certainly oversimplified. Still, a more
realistic picture would share some of the general properties of the
discussed model. Even a gradual transition between the two zones seems
possible. This is true as long as the outer region is capable of
sustaining a self-collimated magnetic field structure through , e.g.,
the Blandford \& Payne mechanism, and the inner region disposes over
enough effective inertia to open up its way along the rotation axis
with large enough Lorentz factors. In particular, the inner region
must not necessarily have negligible angular momentum. In the presence
of magnetic fields, rotation could help to adjust the balance between
the effective inertia and the magnetic confinement.

Within the present modeling, 
our studies show, that the observed opening angle of the M87
jet is best reproduced by models with moderate bulk Lorentz factors,
i.e. $\Gamma \sim 2 - 4$, a range, which is compatible with the
currently favoured values \citep{Biretta+95,CSS04,Biretta+99}.

\begin{acknowledgements}
  Part of this work was supported by the European Community's
  Research Training Network RTN ENIGMA under contract HPRN--CY--2002--00231.
\end{acknowledgements}

\bibliographystyle{aa}
\bibliography{fitM87}

\end{document}